\documentclass[12pt,preprint]{aastex}
\usepackage{graphicx}

\newcommand{\msun}{\rm M$_\odot$}
\newcommand{\rsun}{\rm R$_\odot$}

\shorttitle{A 39~pc double degenerate binary that will merge within 800~Myr} 
\shortauthors{Debes et al.}
\begin{document}
\title{A New Merging Double Degenerate Binary in the Solar Neighborhood\footnote{This paper includes data gathered with the 6.5 meter Magellan telescopes and the 2.5 meter Dupont telescope located at Las Campanas Observatory, Chile.}}
\author{John H. Debes\altaffilmark{1}, Mukremin Kilic\altaffilmark{2}, Pier-Emmanuel Tremblay\altaffilmark{1,3}, Mercedes L{\'o}pez-Morales\altaffilmark{4}, Guillem Anglada-Escude\altaffilmark{5},Ralph Napiwotzki\altaffilmark{6}, David Osip\altaffilmark{7}, Alycia Weinberger\altaffilmark{8}}
\altaffiltext{1}{Space Telescope Science institute, 3700 San Martin Dr., Baltimore, MD 21218, USA}
\altaffiltext{2}{Department of Physics and Astronomy, University of Oklahoma, 440 W. Brooks St., Norman, OK, 73019, USA}
\altaffiltext{3}{Hubble Fellow}
\altaffiltext{4}{Harvard-Smithsonian Center for Astrophysics, 60 Garden Street, Cambridge, MA 01238, USA}
\altaffiltext{5}{Astronomy Unit, School of Mathematical Sciences, Queen Mary, University of London, London E1 4NS, UK}
\altaffiltext{6}{Centre for Astrophysics Research, University of Hertfordshire, Hatfield, AL10 9AB, UK}
\altaffiltext{7}{The Observatories of the Carnegie Institute of Washington, Las Campanas Observatory, Colina El Pino, Casilla 601, La Serena, Chile}
\altaffiltext{8}{Department of Terrestrial Magnetism, Carnegie Institution of Washington, 5241 Broad Branch St., NW, Washington, DC 20015, USA}

\begin{abstract}
Characterizing the local space density of double degenerate binary systems is a complementary approach to broad sky surveys of double degenerates to determine the expected rates of white dwarf binary mergers, in particular those that may evolve into other observable phenomena such as extreme helium stars, Am CVn systems, and supernovae Ia.  However, there have been few such systems detected in local space.  We report here the discovery that WD~1242$-$105, a nearby bright WD, is a double-line spectroscopic binary consisting of two degenerate DA white dwarfs of similar mass and temperature, despite it previously having been spectroscopically characterized as a single degenerate.  Follow-up photometry, spectroscopy, and trigonometric parallax have been obtained in an effort to determine the fundamental parameters of each component of this system.  The binary has a mass ratio of 0.7 and a trigonometric parallax of 25.5~mas, placing it at a distance of 39~pc.  The system's total mass is 0.95~M$_\odot$ and has an orbital period of 2.85 hours, making it the strongest known gravitational wave source ($\log h = -20.78$) in the mHz regime.  Because of its orbital period and total mass, WD~1242$-$105 is predicted to merge via gravitational radiation on a timescale of 740~Myr, which will most likely not result in a catastrophic explosion.
\end{abstract}


\section{Introduction}
As the endpoint of stellar evolution for all stars $<$8~\msun, white dwarfs (WDs) provide key insight into the late evolution of stellar objects.  Double degenerates (DDs), or binary white dwarfs, also open a window into the late evolution of binary systems.  In particular those DDs fated to merge via gravitational wave radiation on timescales shorter than the age of the universe represent a potentially significant progenitor population for Type Ia supernovae (SNe Ia).  Closely orbiting DDs additionally represent an important component to the galactic background gravitational wave radiation \citep{evan87,hils90}.

The first discovery of a close DD system was nearly thirty years ago \citep{saffer88}, with significant efforts to find more via the ESO Supernovae Type Ia Progenitor Survey (SPY) and Extremely Low Mass (ELM) WD surveys \citep{napiwotzki01,napiwotzki03,kilic10,brown10}.  While the SPY survey focused on a large, magnitude limited survey for DDs, the ELM survey has focused entirely on apparently low mass WDs, since these are hard to form without a massive companion \citep{marsh95}.  Both approaches have yielded dozens of new DD systems, which allow for statistical investigations of such binaries.  In this paper, we report the serendipitous discovery that WD~1242$-$105 is a nearby DD system that will merge in a relatively short timescale.

WD~1242$-$105 was first reported as a UV-excess source and misclassified as a subdwarf-B star in the Palomar-Green survey \citep{green86}.  \citet{salim02}, however, identified it as a nearby WD candidate based on its proper motion, which was confirmed spectroscopically by \citet{vennes03}, with further observations that did not reveal it to be of any particular note \citep{salim03,kawka04,kawka06}.  Due to its inferred brightness, gravity, and temperature, it was placed in the local 25~pc sample of white dwarfs \citep{holberg08,giammichele12,sion14}, but lacked any published high resolution optical spectroscopy; presumably it was discovered too late for inclusion in the SPY survey.  For these reasons, we originally targeted this white dwarf in a search of photospheric metal line pollution in nearby white dwarfs \citep{debes10}.  The discovery of a clear secondary component to the spectrum around the H$\alpha$ line led us to further investigate the nature of this system.

In Section \ref{sec:ob} we detail the suite of observations we obtained in order to determine the nature of this binary.  In \S \ref{sec:an} we analyze our results and place strict constraints on the mass and orbital parameters of the binary through simultaneous fitting of synthetic spectroscopy and the relative difference in the components' gravitational redshift.  In \S \ref{sec:disc} we place this system into the context of other double degenerate systems, and present our conclusions in \S \ref{sec:conc}.

\section{Observations}
\label{sec:ob}
\subsection{Magellan MIKE spectra}
We observed WD~1242$-$105 with the blue and red chips of the MIKE spectrograph \citep{bernstein03} installed at the 6.5-m Magellan Clay Telescope at Las Campanas Observatory (LCO; Chile) as part of a survey of nearby WDs for the presence of photospheric metal lines.  All runs used a 0\farcs7$\times$5\farcs0 slit, yielding an average spectral resolution of R $\sim$ 35000 at 3933 \AA. The spectra cover wavelengths between 3335 and 9500~\AA.  Each exposure was taken with a 600~s integration time to ensure sufficient S/N.  Table \ref{tab:vels} lists the 27 observations of the WD, which were taken over multiple epochs starting in March 2008 with the first discovery spectra, and intensive follow-up in April and May 2009.  A nearly full period of the orbit was obtained in May 2009, allowing us to place precise constraints on the orbital period.
 
The data were extracted and flatfielded using the MIKE reduction pipeline written by D. Kelson, with methodology
described in \citet{kelson00} and \citet{kelson03}.  Each spectrum was corrected for heliocentric motion and each epoch was converted to heliocentric Julian date.  The continuum around the H$\alpha$ line was fit with a polynomial and the narrow H-$\alpha$ core was used to measure radial velocities for both components of the binary system via the simultaneous fitting of two Gaussian curves.  In general, this was sufficient to determine the radial velocity of the two components at a precision of $\sim$3-5~km/s.  When the two components of the binary were close to conjunction, the corresponding uncertainty in the line centers increased.  In those cases uncertainties in fitting each velocity were closer to $\sim$20~km s$^{-1}$.

The spectrum was also inspected for any evidence of Ca or Mg absorption, indicative of accretion due to dust or some external source of metal-rich material.  We saw no evidence of this in the raw epoch-to-epoch spectra, but a more detailed analysis is beyond the scope of this paper.  More stringent upper limits for each component of the binary will be presented in a future paper (Todd et al. 2015, in prep.).

\subsection{Time Series Photometry of WD~1242$-$105}
When the first spectra of WD~1242$-$105 showed evidence for the presence of a companion with a very small orbital separation, we conducted a search for eclipses or any sign of photometric variability.  This included observations taken on 2009 May 4 using the MagIC-E2V instrument on the Baade Telescope \citep{eliot}, a fast readout instrument designed for high cadence photometry.  In total we took roughly five hours of data, covering nearly 1.75 orbital periods and using the $V$ filter with exposure times of 60s.   The MagIC instrument had a field of view of 40\arcsec$\times$40\arcsec\, which was large enough to fit both the target and a fainter comparison star.  A photometric aperture of 25 pixels was used for both WD~1242$-$105 and the comparison. 

\subsection{CAPSCAM Astrometry}

We measured the trigonometric parallax using CAPScam \citep{boss09} with the DuPont 2.5m telescope, also at LCO. With a field of view of 6.6\arcmin, the star and a sufficient number of references were observed simultaneously in the standard imaging mode (2048x2048 pixels, with a pixel scale of 0.194\arcsec).
Each observing run consisted of taking 15 to 20 exposures of 30 to 45 seconds depending on the seeing. Five epochs were obtained between June 2009 and July 2010 (8th June, 27th January,  10th April, 22nd June, and 31st July). 
The source extraction, source cross matching, geometric calibration and astrometric solution have been obtained using the ATPa software \citep{anglada12b,boss09}. The overall precision for the target star and the reference frame stars is 1 mas/epoch. The reference stars are selected by the software iteratively based on their epoch to epoch RMS. A robust reference frame of 33 stars is used. Parallax and proper motion of the reference stars is also obtained as a by-product. Several reference stars have unambiguous USNO B1 and 2MASS counter parts and the B-K color with the 2MASS magnitudes are used to estimate the photometric distance to them. Reference stars with parallaxes $<$5~mas were used, taken either from direct measurements or from photometric estimates. At the end the correction from relative to absolute parallax/proper motion is obtained based on 13 reference stars. The final parallax and corresponding distance estimation are 25.5$\pm$0.9~mas and 39.2$^{+1.4}_{1.3}$~pc.  The statistical uncertainty in the parallax is obtained from by a Monte-Carlo resampling of the astrometry at the same observation dates which properly takes into account all significant parameter correlations.  The original estimated spectroscopic distance to WD~1242$-$105 was 25~pc when it was believed to be a single WD--extrapolating the spectroscopic distance for two WDs results in a distance of 35~pc, consistent with our parallax measurement.

\section{Analysis}
\label{sec:an}
In this Section, we describe our analysis of the various observations in order to better constrain the properties and fate of the WD~1242$-$105 system.  The combination of all our constraints allow for a determination of fundamental parameters for both WDs, which in turn allows us to determine whether the system will merge and if so, whether it is massive enough to become a progenitor of a Type Ia supernova.

\subsection{Ephemeris Determination}

Table \ref{tab:vels} lists the measured velocities for all epochs of the WD~1242$-$105 system.  The velocities were fit with the non-linear least squares fitting routine {\em curvefit.pro} in the Interactive Data Language, using sinusoidal curves of the form $K \cos(p HJD-HJD_{\rm o})+\gamma$, with $p$ equal to the period, $\gamma$ equal to the velocity offset of each component, $HJD_{\rm{o}}$ equal to the reference epoch, and $K$ equal to the velocity semi-amplitude.  The long baseline of observations as well as the dense sampling of the orbit in May 2009 allowed the fitting routines to converge on a precise period for the components.  These fits give the final ephemeris of the system:
% period=0.11876571+/- 0.000002
% t0=2454970.5901+/-0.0001 
%need to check measurement of K2, night3
\begin{equation}
\phi= (HJD- HJD_{\rm o})p
\end{equation}

where $HJD_{\rm o}$=2454970.5901$\pm$0.0001, and $p$=0.118765$\pm$0.000002).

\begin{figure}
\plotone{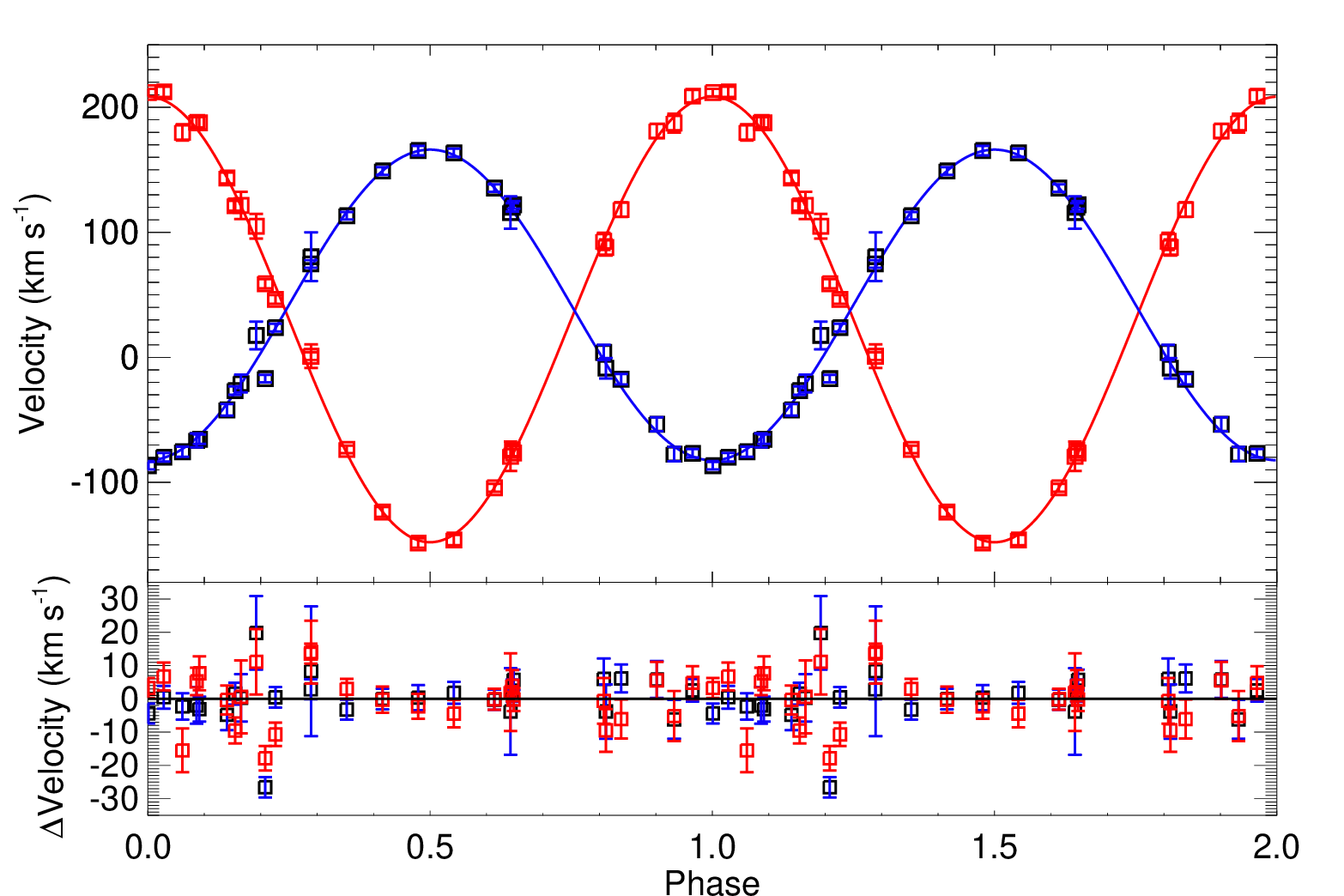}
\caption{\label{fig:f1} Phased radial velocity curve of the two components of the WD~1242$-$105 DD system.  The red points are derived from Gaussian fits to the core of the H$\alpha$ feature for the less massive component, while the blue points are for the more massive component.  The solid lines correspond to the best fit orbit for both components, with residuals plotted in the lower panel.}
\end{figure}

Figure \ref{fig:f1} shows the final phased radial velocities for both components, while Table \ref{tab:orbit} details the velocity semi-amplitudes and $\gamma$ for each component of the system as well as the inferred mass functions for each component.  Because of the two components' differing masses, their velocity offsets represent a combination of the binary's systemic velocity and the gravitational redshift of each component.  This can be used in combination with the traditional mass-radius relations to solve uniquely for the masses of the two systems, which we investigate in \S \ref{s:mass}.

\subsection{Photometric Variability}

\begin{figure}
\plottwo{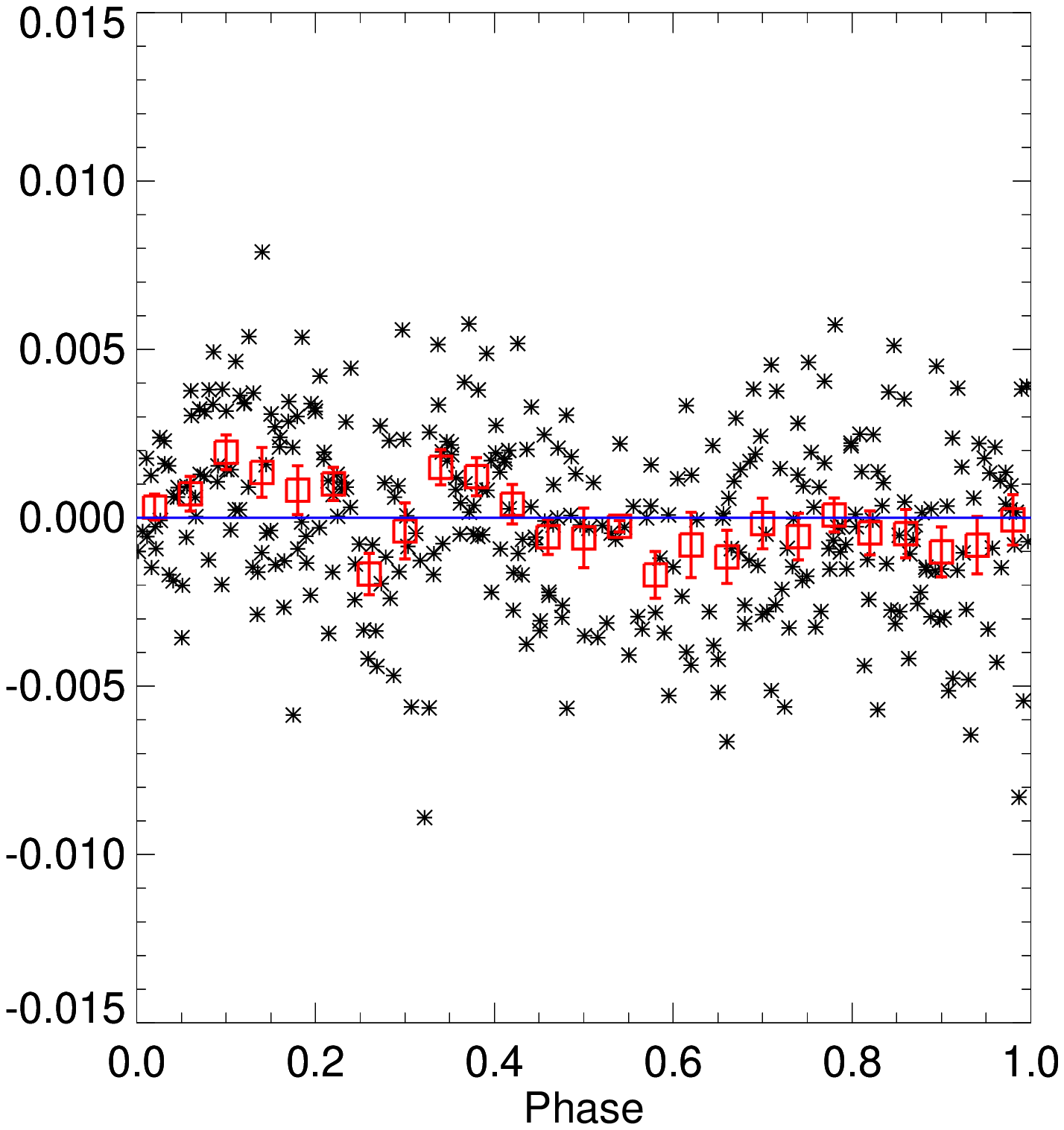}{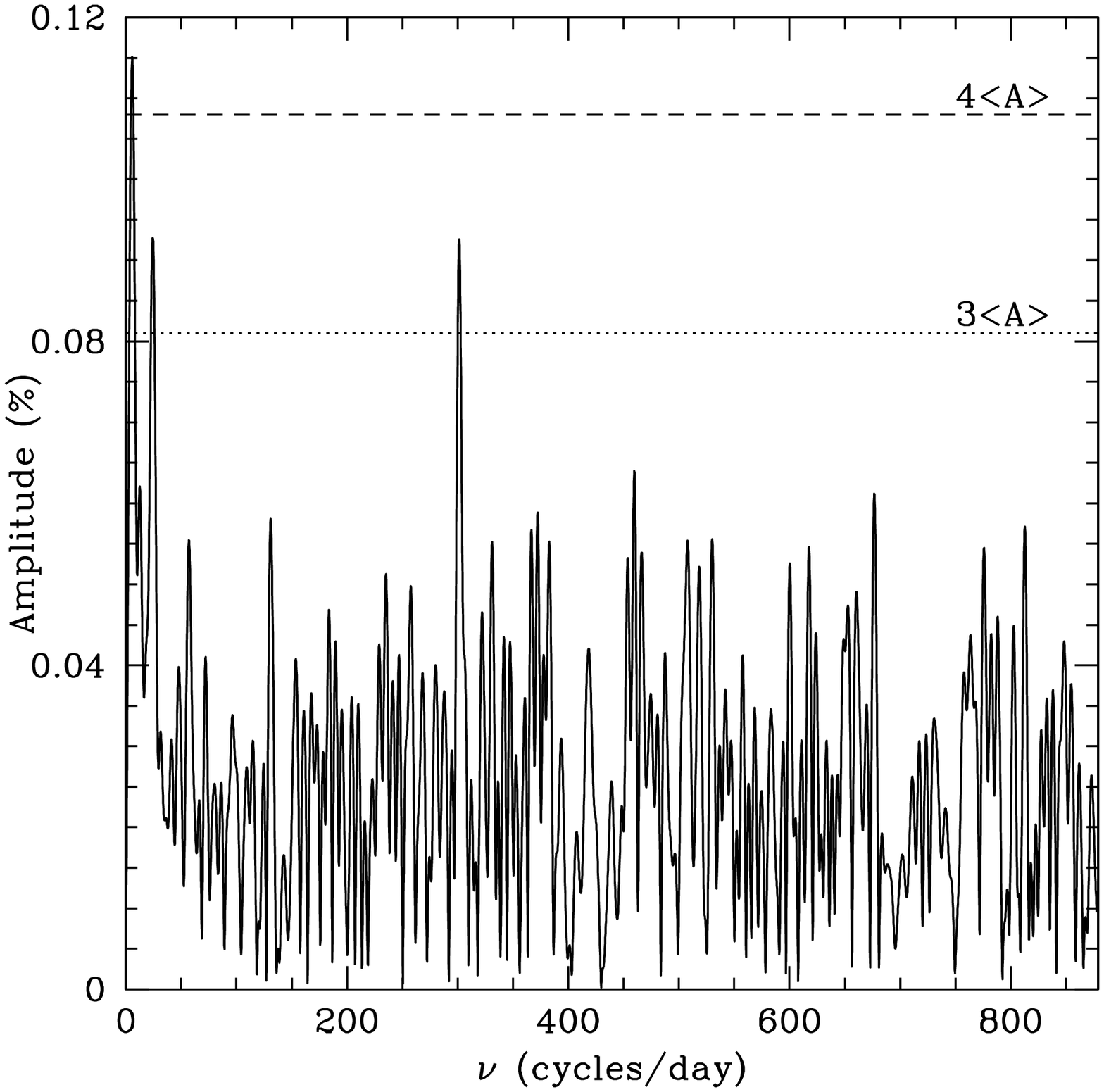}
\caption{\label{fig:f2} (left) MagIC-E2V high cadence photometry of WD~1242$-$105 phased to its orbital period.  Black points show the raw 60s exposures with roughly 3mmag precision.  Overplotted in red points are the phase-binned photometry with 9 minute sampling and an rms of less than 1mmag.   (right) Fast fourier transform of the data, showing no significant peaks.  A long period peak is marginally significant, but does not match the period of the binary.  We attribute this signal to slow changes in atmospheric conditions.}

\end{figure}

Figure \ref{fig:f2} shows both the phase folded light-curve of the V band photometry of WD~1242$-$105, with a measured rms of 3 milli-mags and 60-s sampling, along with a Fourier transform (FT) of the photometry.  No obvious periodicity is seen in the data to a 4-$\sigma$ level of 1.1 mmag, and the overall standard deviation of the photometry matches the estimates of the photometric uncertainty.  A peak in the FT at 4.28 h is seen, but since this is not coincident with the orbit of the system we attribute this to slowly varying atmospheric extinction.  The comparison star most likely had a different SED than WD~1242$-$105, and would suffer from differential atmospheric extinction which could explain the small variation.  Rebinning the data along the phase of the orbit does not show any obvious additional structure, implying that the two components are well detached, non-eclipsing, and not suffering from any tidal distortion.

\section{Determination of WD properties}
\label{s:mass}
From the orbital radial velocities, spectroscopy, parallax, and photometry we have several independent constraints on the mass and radius of each component of the WD~1242$-$105 system.  We calculate the mass of each component two ways: via the difference in the gravitational redshifts of each component, and by simultaneous fitting of the spectra using the observed mass ratio and parallax as additional constraints.

\subsection{The mass ratio and gravitational redshift of WD~1242$-$105}
\label{sec:grav}
We can calculate the mass ratio of the two components $q=0.70\pm0.01$ from the ratio of the radial velocity semi-amplitudes.  The difference in systemic velocity, or $\gamma$ provides a constraint on the difference of the two objects' gravitational redshifts:

\begin{equation}
\label{eq:grav}
\Delta\gamma=\frac{G}{c} \left(\frac{M_1}{R_1}-\frac{M_1}{qR_2} \right)
\end{equation}

where $M_1$ and $R_1$ denote the mass and radius of the more massive companion respectively, and $R_2$ denotes the radius of the less massive companion.  We measure a velocity offset difference of 11.6$\pm$1.3~km~s$^{-1}$.  Under the assumption of theoretical mass-radius relations for white dwarfs (which does require an estimate of T$_{\rm eff}$), one can uniquely fit the mass ratio and gravitational redshift to give masses of the two components \citep{napiwotzki02}.  In a recent study by \citet{holberg12}, most field DAs showed good agreement with theoretical mass-radii relations.  Similarly, careful observations of double degenerates show that low mass WDs that have experienced post-common envelope evolution also generally behave consistently with expected model mass-radius relations \citep{bours14}.  To obtain the mass estimates, we minimized a $\chi^2$ metric for the expected $\Delta\gamma$ for a given primary mass and mass ratio, and using the derived T$_{\rm eff}$ from \S \ref{sec:synth}.  Using this approach, we find masses of 0.56$^{+0.05}_{-0.07}$ and 0.39$^{+0.04}_{-0.05}$ M$_{\odot}$.  We compare this determination of the masses with those determined in the next section.

\subsection{Simultaneous Spectroscopic and Photometric spectral energy distribution modeling}
\label{sec:synth}
Barring a measurement of the inclination of the binary orbit, the unknown masses and radii of the binary components need to be disentangled with additional information that can be derived from modeling the two components' optical spectra and their photometric spectral energy distributions (SED).  This modeling, in concert with the constraints derived from the mass ratio and the parallax allows determination of the WD fundamental parameters complementary to \S \ref{sec:grav}.

In an era of all-sky surveys in the UV through mid-IR, high quality SEDs are now routine.  In particular, WD~1242$-$105 is within the sky coverage of GALEX GR7 \citep{martin05}, SDSS DR9 \citep{ahn12}, 2MASS \citep{skrutskie06}, and ALLWISE \citep{wright10} , resulting in 14 photometric measures of the system's SED.
With an accurate measure of the systemÕs parallax, we determine both
$T_{\rm eff}$ and $\log g$ by fitting synthetic spectra to the observed
photometry and spectroscopy under the constraints of the observed mass
ratio. Fitting the spectroscopic and photometric measurements alone
introduces degeneracies where multiple similar temperatures and
gravities are possible.

The procedure to fit the spectrum is the same
as that used for single WDs where the profiles of the hydrogen Balmer
lines are compared to detailed model atmospheres
\citep{bergeron92,liebert05}. We rely on a combination of spectra
taken when the components were well separated in velocity space, namely the spectra obtained on HJD 2454548 (See
Table \ref{tab:vels}). Each individual (blended) line was normalized to a continuum
set to unity at a fixed distance from the line center, for both
observed and model spectra. The atmospheric parameters are then found
using the nonlinear least-square method of Levenberg-Marquardt
\citep{press86}. The uncertainties on fitted values were derived from a combination of the covariance matrix of the spectroscopic fitting algorithm, which mostly impacts $T_{\rm eff}$, and error propagation of the trigonometric parallax and mass ratio uncertainties, which mostly influence $\log g$ values.  
We determine $T_{\rm eff}$ for both components from
the spectroscopic fit but the $\log g$ values are fixed from the
result of the photometric fit.

For our procedure of fitting the photometry, we included the Sloan $ugriz$ photometry and converted
these measurements into flux densities using the
appropriate filters, which are then compared with the predictions from
model atmosphere calculations \citep{bergeron97,holberg06}. We apply a
correction to the $u$, $i$, and $z$ bands of $-$0.040, +0.015, and
+0.030, respectively, to account for the offsets between the SDSS filter zeropoints
and the AB magnitude system \citep{eisenstein06}. From the photometry
we only fit the solid angle with the constraint from the trigonometric
parallax. From the observed mass ratio and spectroscopic $T_{\rm eff}$
values, we can then determine both gravities assuming a mass-radius
relation. We iterate on the spectrosopic and photometric fits until we
converge to a solution on the atmospheric parameters.

To fit the observations, we rely directly on a grid of mean 3D spectra
from pure-hydrogen atmosphere 3D simulations \citep{tremblay13}. In
this range of $T_{\rm eff}$, 3D effects on the gravities can be quite
dramatic. The gravities were converted to masses and radii using
evolution sequences with thick hydrogen layers from \citet{fontaine01}
for the C/O core component and \citet{althaus01} for the lower mass He
core component.  The choice of composition for the core comes from the implied masses determined in \S \ref{sec:grav}, He-core WDs generally experienced extreme mass loss during the RGB, leaving a core less than 0.5~\msun.

\begin{figure}
\plotone{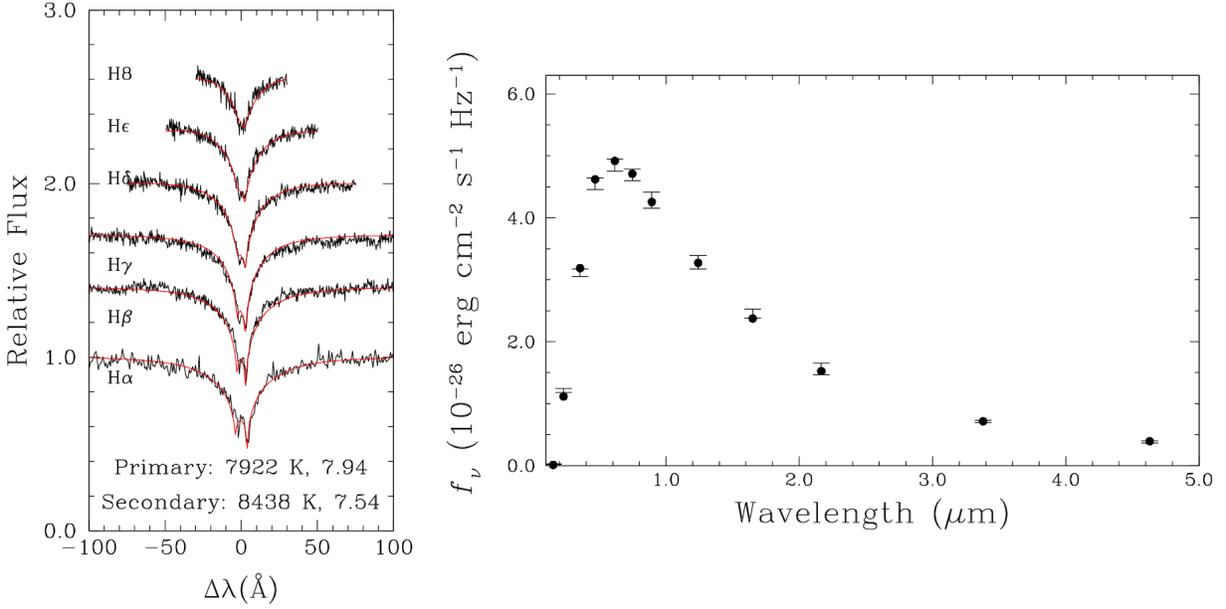}
\caption{\label{fig:f3}   (left) Comparison between synthetic spectra of WD~1242$-$105 and observed spectroscopy of the system.  Spectral lines from the primary are blue shifted in the spectrum, while the secondary's spectral lines are redshifted. The Balmer series of hydrogen lines are compared to simultaneous model fits of each component of the DD system.  Black lines are the observed MIKE spectra, while red are the model spectra. (right) Comparison of observed UV through Mid-IR photometry of the WD~1242$-$105 system (black error bars) compared to the models (black circles).  The parallax and mass ratio were used to simultaneously fit both components to the spectroscopy and photometry.}
\end{figure}

Figure \ref{fig:f3} shows the resulting model fits to both the optical spectroscopy and the SED of the two components under the assumption of the parallax and mass ratio for the system.  In addition to a good fit to the Sloan photometry, the predicted photometry for GALEX, 2MASS, and ALLWISE photometry is consistent within the uncertainties.  Table \ref{tab:orbit} lists the final $T_{\rm eff}$, $\log~g$, and masses as derived from our fitting procedure.  From this procedure we obtain exact agreement in the masses of the two components with that determined in \S \ref{sec:grav}.  This is not completely surprising as the two measurements are linked via the inferred mass ratio of the binary system as well as the same theoretical WD mass-radius relationships, but does provide confidence in the resulting answer.  For our further discussions, we adopt the masses and uncertainties derived from the spectroscopic and photometric fitting.

In summary, the WD~1242$-$105 double degenerate binary is composed of a C/O core WD with a mass of 0.56 \msun\ and a less massive 0.39 \msun\ Helium core WD.  They orbit each other in a period of less than 2.85 hours, with an inclination to the line-of-sight of 45.1$^\circ$.  The semi-major axis of their orbit is 1~\rsun, which at a distance of 39.2~pc corresponds to a maximum angular separation of 120~$\mu$as.

\section{WD 1242$-$105: The Past, Present, and Future}
\label{sec:disc}
The orbital periods of Post-Common-Envelope-Binaries (PCEBs) containing
low-mass He-core WDs tend to be shorter than the orbital periods of PCEBs
containing more massive C/O core WDs \citep{zorotovic11}. The reason for
this trend is that shorter period systems interact earlier in their evolution
and experience enhanced mass loss, ending up as lower mass He-core WDs.
With an orbital period of 0.118765 d and a 0.39 $M_{\odot}$ binary member,
WD~1242$-$105 follows this trend.

Recently, many close binary white dwarfs with low mass progenitors have been discovered along with a significant number of longer period and more massive binary systems \citep{nelemens05,kilic12}, with a large number being future merger products.  WD~1242$-$105 represents an interesting case as it is close (d$\ll$100~pc), with a relatively short period and a mass ratio close to 1, but also with a fairly high total mass to the system compared to other discoveries.  It is a complementary detection to both the SPY survey \citep{napiwotzki01} and the ELM survey \citep{brown10}. 

\begin{figure}
\plotone{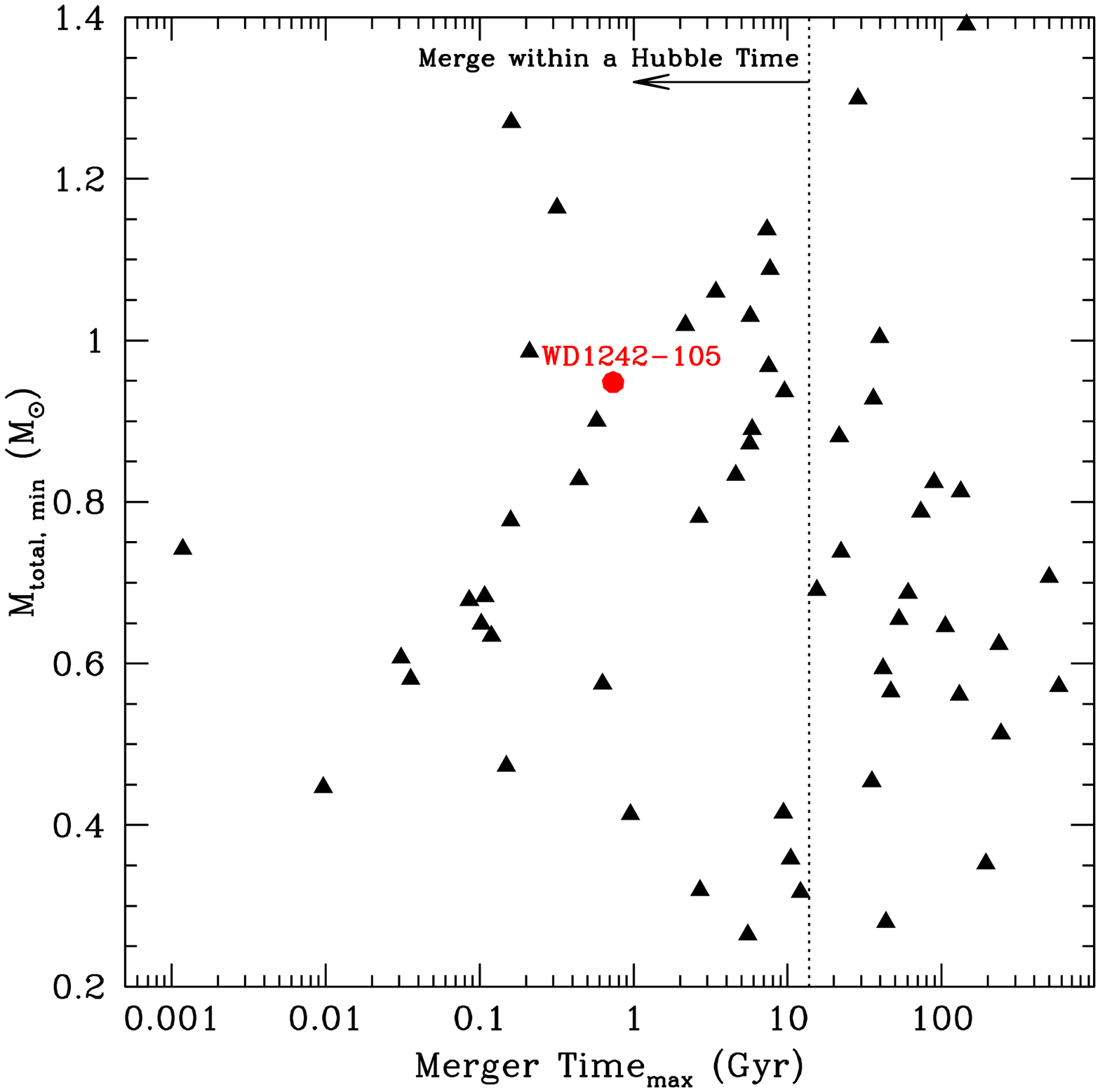}
\caption{\label{fig:f4} Gravitational wave merger time vs. total system mass for
double white dwarfs in the ELM Survey \citep{brown13,kilic14,gianninas14}
and WD 1242$-$105 (red filled circle in online version). For the ELM Survey sample, we plot the minimum total
system mass (assuming $i=90^{\circ}$) when the orbital inclination is unknown,
and the correct system mass when the inclination is known either from eclipses
or ellipsoidal variations. Objects to the left of the dotted line will merge within a Hubble time.}
\end{figure}

Figure \ref{fig:f4} shows the total system mass and merger time
for double white dwarfs in the ELM Survey and WD~1242$-$105. The latter has
accurate mass measurements for both components. We plot the
minimum total system mass (and hence the maximum gravitational wave merger time)
for the ELM white dwarf sample, unless the orbital inclination is known from
eclipses or ellipsoidal variations. With a merger time of 737~Myr and total mass
of 0.95~$M_{\odot}$, WD~1242$-$105 is one of the nearest, most massive, and quickest merger
systems known. 

There are other double degenerate systems that are likely to be closer than WD~1242$-$105, but lack measured parallaxes or will not merge within a Hubble time.  
Another merging WD system with a period roughly twice as long as WD~1242$-105$, NLTT 53177 \citep{karl03}, may be closer by a few parsecs, given the inferred spectroscopic distance of its two components.  WD~1242$-$105 is a near twin of the compact component to the WD~1704+481 system \citep{maxted00b}, which consists of three white dwarfs, two of which are in an orbit with a period of 0.145~d.  The mass ratio of this pair is also 0.7, with a similar difference in their gravitational redshifts.  The spectroscopic distance of the distant third component is 40~pc \citep{gianinas11}, which is similar to WD~1242$-105$'s parallax.
Finally, there are other double degenerate systems within 25 pc of the Sun
\citep{holberg08}, but those have gravitational wave merger times longer than a Hubble time.

We can also investigate the eventual fate of the system. Figure 5 shows WD~1242$-$105 
compared to other DDs and relative to the stability criteria of \citet{marsh04}, which dictates whether objects merge violently with the possibility of detonation or stably through Roche lobe overflow mass transfer. 
All of the massive merger systems in the ELM Survey are found due to the
$\approx0.2 M_{\odot}$ ELM white dwarfs. Hence, these tend to have extreme
mass ratios ($q\approx0.2$), which should lead to stable mass transfer
\citep{marsh04} AM CVn objects. On the other hand, WD 1242$-$105
has $q=0.7$, which will lead to unstable mass transfer and a merger.

Simulations of moderately massive C/O core WDs with thick He layers (the case for WD~1242$-$105 when
it merges) shows that a cataclysmic
explosion could occur and result in a SNe Ia like phenomena \citep{sim12,shen10} due to
the detonation of the helium-shell or double detonation of both the helium layer and the C/O WD.
\citet{dan14} performed a large parameter space exploration of the merger products for
CO+CO and CO+He WDs. They find that if the timescale for triple-$\alpha$ reactions is less than
the dynamical timescale ($\tau_{\rm nuc} \leq \tau_{\rm dyn}$), a helium-shell detonation would occur
\citep[see also][]{guillochon10}.
Their simulation involving a 0.4 $M_{\odot}$ He-core WD with a 0.55 $M_{\odot}$ CO WD (similar
to WD~1242$-$105) reaches a maximum temperature of $\log{T_{\rm max}} = 8.4$ K and density
$\log{\rho_{T_{\rm max}}} = 4.65$ g cm$^{-3}$. This model has $\tau_{\rm nuc} \gg \tau_{\rm dyn}$, making detonation unlikely. The merger will most likely leave behind an extreme helium
star (R Cor Bor) with a mass close to 0.9 $M_{\odot}$ \citep{saio00}.

WD~1242$-$105 is also a signiÞcant source of gravitational waves in the mHz frequency range.
At a distance of 39 pc and $i = 45.1^{\circ}$, we expect the gravitational wave strain at Earth
$\log h = -20.78$ at $\log \nu$ (Hz) = $-3.71$ \citep{roelofs07}.
 Unfortunately, this places WD~1242$-$105 outside of the expected sensitivity of the eLisa mission \citep{amaro12}.  However, of known ELM systems, it is the strongest source of gravitational wave radiation at mHz frequencies, primarily because of its proximity to Earth.

Finally, we speculate on the discovery of WD~1242$-$105 and the prospects for finding more systems like it within local space.  Roughly 5\%, and as many as 13\% of white dwarfs are in close binaries, if one assumes binomial probabilities based on the detection of 2 short period DDs within a sample of 44 \citep{maxted99}.  The SPY survey of DA WDs found 39 DDs among 679 observed WDs, implying again a 5.7\% frequency \citep{koester09}. The local sample of WDs within 20~pc ($\sim$126) has 4 reported instances of unresolved DDs \citep{holberg08}, implying a frequency of $\sim$3\% but no more than 7\%.  These numbers are broadly consistent with each other, however there could be at least three more local WDs that are actually undiscovered DDs.  Given that the 40~pc WD sample should include $\sim$70 or so close DD systems, many local DDs are still unaccounted for, but should be apparent with the launch of GAIA--these systems will appear over-luminous for their given composite gravities, as WD~1242$-$105 was.  Moderate resolution optical echelle spectroscopy of WDs with 10-20 minute cadences, such as what was obtained for WD~1242$-$105, are sufficient to detect DDs with short periods even with cooler $T_{\rm eff}$.  A volume limited survey of double degenerates would provide tight constraints on the degenerate population of binaries that may participate in mergers and cataclysmic explosions.

 \section{Conclusion}
 \label{sec:conc}
 We have detected a new nearby merging double white dwarf binary system, WD~1242$-$105, previously believed to be a single WD located within 25~pc from Earth.  Our radial velocity measurements, photometry, and astrometry show it to be a pair of white dwarfs at a distance of 39.2~pc, with a period of 0.1187~d, and possessing a mass ratio of 0.7.  The total mass of the system is 0.95~\msun, and since the two components are hydrogen-rich and of similar luminosity we can determine the difference in their gravitational redshifts and thus their individual masses.  We also simultaneously fit photometry and spectra of the system to calculate the individual masses of the binary a second way, which agrees to within the uncertainties.  The short orbital period of the binary guarantees that it will merge within 1~Gyr, possibly in the form of an under-luminous supernova or extreme helium star and makes it one of the strongest known gravitational wave sources in the mHz regime.  Regardless of its eventual fate, WD~1242$-$105 represents an interesting example of a merging DD system that is bright and close to the Earth.
 
 \acknowledgements
The authors would like to thank the numerous support staff and scientists at the Las Campanas Observatory, who make world-class observing routine by their high level of help and knowledge.  This research has made use of the VizieR catalogue access tool, CDS, Strasbourg, France.  MK gratefully acknowledges the support of the NSF and NASA under grants AST-1312678 and NNX14AF65G, respectively.  JHD acknowledges support through a contract to the Association of Universities for Research in Astronomy via the European Space Agency.  P.-E. T is supported by NASA through Hubble Fellowship grant \#HF-51329.01 awarded by the Space Telescope Science Institute.
 
 {\it Facilities:} \facility{Magellan:Clay (MIKE)}, \facility{Magellan:Baade (MagIC)}, \facility{Dupont (CAPSCAM)}


\begin{figure}
\plotone{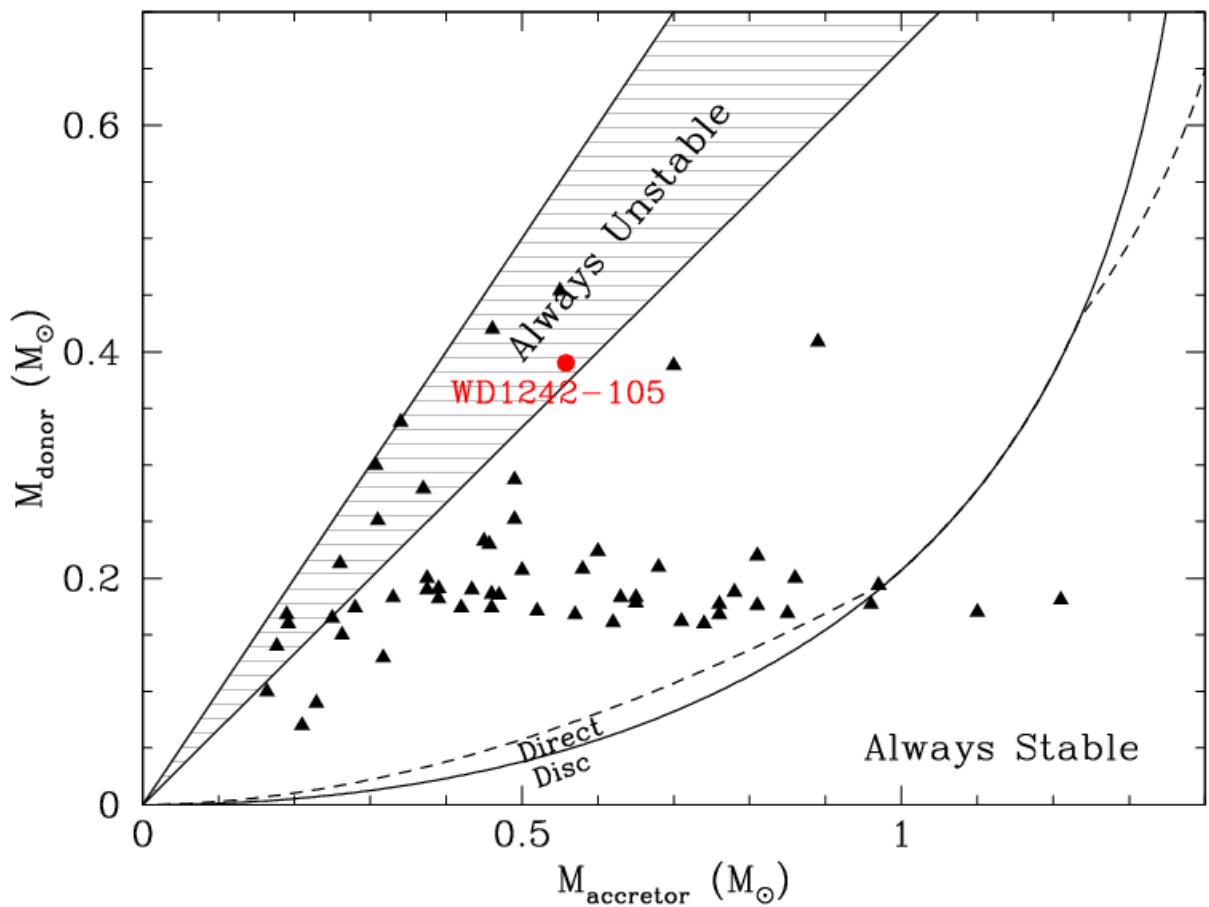}
\caption{\label{fig:f5} Plot of mass ratio of the same systems shown in Figure \ref{fig:f3},
including WD1242, compared to the stability criteria of \citet{marsh04}.  With masses of 0.39 and 0.56 \msun\ and unstable transfer leading to a merger, the resulting merged remnant will likely be an extreme helium star, but could result in a sub-Chandrasekhar supernova.}
\end{figure}

\begin{deluxetable}{cccccc}
\tablecolumns{3}
\tablewidth{0pt}
  \tablecaption{Measured Radial Velocities of WD~1242$-$105 \label{tab:vels}}
  
\tablehead{  
\colhead{UT Date} & \colhead{HJD-240000} & \colhead{K$_1$} & \colhead{$\sigma$ K$_1$} & \colhead{K$_2$} & \colhead{$\sigma$ K$_2$} \\
& & \colhead{km s$^{-1}$} & \colhead{km s$^{-1}$}  & \colhead{km s$^{-1}$} &  \colhead{km s$^{-1}$}
	}
\startdata

2008 Mar 23  &               54548.72600  &      -77 &        5 &      187 &        7 \\
  &                                   54548.73416  &      -86 &        3 &      211 &        3\\
 &                                    54548.74133  &      -75 &        3 &      179 &        6 \\
2009 Apr 16 &                54937.55058  &       -8 &        8 &       87 &        6\\
 &                                   54937.64928&      115 &       13  &      -79 &       11 \\
 &                                   54937.71134  &      -20 &        7 &      121 &       10\\
2009 May 10 &              54962.47196 &      121 &        3 &      -76 &        3 \\
2009 May 11 &              54962.53042 &      -41 &        4 &      143 &        4 \\
 &                                  54962.53848 &      -17 &        3  &       58 &        3 \\
 &                                  54962.70545  &      135 &        3 &     -104 &        3 \\
 &                                  54962.77412  &       17 &       11 &      104 &        9\\
2009 May 17 &             54969.49811  &        3 &        6 &       92 &        6\\
2009 May 18 &             54969.53122 &      -66 &        5 &      188 &        4 \\
 &                                  54969.59765 &      120 &        3 &      -74 &        4 \\
 &                                  54969.67401  &        0 &        3 &       74 &        3\\
 &                                  54970.57065  &      118 &        5 &      -17 &        4 \\
 &                                 54970.57816  &      -53 &        5 &      181 &        5\\
 &                                 54970.58567  &      -76 &        3 &      208 &        4 \\
 &                                 54970.59318  &      -79 &        3 &      212 &        4\\
 &                                 54970.60070 &      -65 &        3 &      187 &        5 \\
 &                                 54970.60821 &      -26 &        3 &      121 &        3 \\
 &                                 54970.61669  &       23 &        3 &       46 &        3\\
 &                                 54970.62419  &       80 &       19 &        0 &        9 \\
 &                                 54970.63170 &      113 &        3  &      -73 &        3 \\
 &                                 54970.63921  &      149 &        3 &     -123 &        3\\
 &                                 54970.64672  &      165 &        3 &     -148 &        3\\
 &                                 54970.65423 &      163 &        3 &     -146 &        4  \\
\enddata
\end{deluxetable}

\begin{deluxetable}{ccccc}
\tablecolumns{5}
\tablewidth{0pt}
  \tablecaption{The WD~1242$-$105 binary \label{tab:orbit}}
  
\tablehead{ 
\colhead{Parameter} & \colhead{Primary} & \colhead{Uncertainty} & \colhead{Secondary} & \colhead{Uncertainty}
}
\startdata
K (km s$^{-1}$) & 124 & 1.2 & 178 & 1.4 \\
$\gamma$ (km s$^{-1}$) & 41.9 & 0.8 & 30.3 & 1.0 \\
%a (\rsun) & 
%a ($\mu$as) & & & & \\
M (M$_\odot$) & 0.56 & 0.03 & 0.39 & 0.02 \\
$T_{\rm eff}$ (K) & 7935 & 92 & 8434 & 36 \\
log $g$ & 7.94 & 0.05 & 7.54 & 0.05 \\
T$_{\rm cool}$ (Gyr) & 1 & ... & 0.6 & ... \\
\enddata
\end{deluxetable}

\end{document}